\documentstyle[epsf,twocolumn,prl,aps]{revtex}
\begin{document}

\draft
%
%
\newcommand{\nc}{\newcommand}
\nc{\bea}{\begin{eqnarray}}
\nc{\eea}{\end{eqnarray}}
\nc{\beq}{\begin{equation}}
\nc{\eeq}{\end{equation}}
\nc{\bi}{\begin{itemize}}
\nc{\ei}{\end{itemize}}
\nc{\la}[1]{\label{#1}}
\nc{\half}{\frac{1}{2}}
\nc{\p}{{\rm p}}
\nc{\barp}{\bar{\rm p}}
\nc{\EH}{{}^3{\rm H}}
\nc{\EHe}{{}^3{\rm He}}
\nc{\UHe}{{}^4{\rm He}}
\nc{\ZLi}{{}^7{\rm Li}}
\nc{\DH}{{\rm D}/{\rm H}}
\nc{\ZLiH}{{}^7{\rm Li}/{\rm H}}
\nc{\et}{\eta_{10}}
\nc{\fsky}{f_{\rm sky}}
\nc{\fwhm}{\theta_{\rm fwhm}}
\nc{\fwhmc}{\theta_{{\rm fwhm},c}}
\nc{\GeV}{\mbox{ GeV}}
\nc{\MeV}{\mbox{ MeV}}
\nc{\keV}{\mbox{ keV}}
\nc{\etal}{{\it et al.}}
\nc{\x}[1]{}
%
%
\nc{\AJ}[3]{{Astron.~J.\ }{{\bf #1}{, #2}{ (#3)}}}
\nc{\anap}[3]{{Astron.\ Astrophys.\ }{{\bf #1}{, #2}{ (#3)}}}
\nc{\ApJ}[3]{{Astrophys.~J.\ }{{\bf #1}{, #2}{ (#3)}}}
\nc{\apjl}[3]{{Astrophys.~J.\ Lett.\ }{{\bf #1}{, #2}{ (#3)}}}
\nc{\app}[3]{{Astropart.\ Phys.\ }{{\bf #1}{, #2}{ (#3)}}}
\nc{\araa}[3]{{Ann.\ Rev.\ Astron.\ Astrophys.\ }{{\bf #1}{, #2}{ (#3)}}}
\nc{\arns}[3]{{Ann.\ Rev.\ Nucl.\ Sci.\ }{{\bf #1}{, #2}{ (#3)}}}
\nc{\arnps}[3]{{Ann.\ Rev.\ Nucl.\ and Part.\ Sci.\ }{{\bf #1}{, #2}{ (#3)}}}
\nc{\MNRAS}[3]{{Mon.\ Not.\ R.\ Astron.\ Soc.\ }{{\bf #1}{, #2}{ (#3)}}}
\nc{\mpl}[3]{{Mod.\ Phys.\ Lett.\ }{{\bf #1}{, #2}{ (#3)}}}
\nc{\Nat}[3]{{Nature }{{\bf #1}{, #2}{ (#3)}}}
\nc{\ncim}[3]{{Nuovo Cimento }{{\bf #1}{, #2}{ (#3)}}}
\nc{\nast}[3]{{New Astronomy }{{\bf #1}{, #2}{ (#3)}}}
\nc{\np}[3]{{Nucl.\ Phys.\ }{{\bf #1}{, #2}{ (#3)}}}
\nc{\pr}[3]{{Phys.\ Rev.\ }{{\bf #1}{, #2}{ (#3)}}}
\nc{\PRC}[3]{{Phys.\ Rev.\ C\ }{{\bf #1}{, #2}{ (#3)}}}
\nc{\PRD}[3]{{Phys.\ Rev.\ D\ }{{\bf #1}{, #2}{ (#3)}}}
\nc{\PRL}[3]{{Phys.\ Rev.\ Lett.\ }{{\bf #1}{, #2}{ (#3)}}}
\nc{\PL}[3]{{Phys.\ Lett.\ }{{\bf #1}{, #2}{ (#3)}}}
\nc{\prep}[3]{{Phys.\ Rep.\ }{{\bf #1}{, #2}{ (#3)}}}
\nc{\RMP}[3]{{Rev.\ Mod.\ Phys.\ }{{\bf #1}{, #2}{ (#3)}}}
\nc{\rpp}[3]{{Rep.\ Prog.\ Phys.\ }{{\bf #1}{, #2}{ (#3)}}}
\nc{\ibid}[3]{{\it ibid.\ }{{\bf #1}{, #2}{ (#3)}}}


\wideabs{
\title{Constraining Antimatter Domains in the Early Universe with
Big Bang Nucleosynthesis}

\author{Hannu Kurki-Suonio\cite{mailh} and Elina Sihvola\cite{maile}}
\address{Helsinki Institute of Physics and Department of Physics,
         P.O.Box 9, FIN-00014 University of Helsinki, Finland}


\maketitle

\begin{abstract}
We consider the effect of a small-scale matter-antimatter domain structure
on big bang nucleosynthesis and place upper limits
on the amount of antimatter in the early universe.  For small domains,
which annihilate before nucleosynthesis, this limit comes
from underproduction of $\UHe$.  For larger domains, the limit comes
from $\EHe$ overproduction.  Most of the $\EHe$ from $\bar{\rm p}\UHe$
annihilation is annihilated also.  The main source of $\EHe$
is photodisintegration of $\UHe$ by the electromagnetic cascades
initiated by the annihilation.

\end{abstract}

}

%
%

If the early universe was homogeneous, antimatter annihilated
during the first millisecond. 
However, baryogenesis could have been
inhomogeneous, possibly resulting in a negative net baryon number
density in some regions\cite{baryogen,GS98}.
After local annihilation these regions would have only
antimatter left, resulting in a matter-antimatter domain structure.

There are many proposed mechanisms for baryogenesis\cite{baryogen}.
In models connected with inflation, 
there is no a priori constraint on the distance scale of the 
matter-antimatter domain structure that may be generated.
If the distance scale is small, the antimatter domains would have
annihilated in the early universe, and the presence of matter today
indicates that originally there was less antimatter than matter.

We consider here such a scenario: a baryoasymmetric universe where the
early universe contains a small amount of antimatter in the
form of antimatter domains surrounded by matter.  We are interested
in the effect of antimatter
on big bang nucleosynthesis (BBN)\cite{BBNobs}.
Much of the earlier work
on antimatter and 
BBN\cite{Steigman,SymmABBN,Aly,He3ABBN,InjABBN,RJ98,AF98}
has focused on a baryon-antibaryon symmetric
cosmology\cite{SymmABBN,Aly} or on homogeneous injection of antimatter
through some decay process\cite{InjABBN}.

The smaller the size of the antimatter domains, the earlier they
annihilate.  Domains smaller than 100 m at 1 MeV, corresponding to 
$2\times10^{-5}$ pc today, would annihilate
well before nucleosynthesis and would leave no observable remnant. 

The energy released in annihilation thermalizes with the
ambient plasma and the background radiation, if the energy release
occurs at $T > 1$ keV.  If the annihilation occurs later, Compton
scattering between heated electrons and the
background photons transfers energy to the microwave background, but is
not able to fully thermalize this energy. 
The lack of observed distortion in the cosmic microwave
background (CMB) spectrum
constrains the energy release occuring after $T = 1$ keV to below
$6\times10^{-5}$ of the CMB energy\cite{Wright94,Fixsen96}.
This leads to progressively stronger
constraints on the amount of antimatter annihilating at later times,
as the ratio of matter and CMB energy density is getting larger.
Above $T \sim 0.1$ eV the baryonic matter energy density is smaller
than the CMB
energy density, so the limits on the antimatter fraction annihilating
then are weaker than $6\times10^{-5}$.

For scales larger than  10 pc (or $10^{11}$ m at $T = 1$ keV)
the tightest
constraints on the amount of antimatter come from the CMB spectral
distortion, and for even larger scales from the 
cosmic diffuse gamma spectrum\cite{CRG98}.

We consider here intermediate domain sizes, where most of the
annihilation occurs shortly before, during, or shortly after
nucleosynthesis, at temperatures between 1 MeV and 10 eV.
The strongest constraints on the amount of antimatter at these distance
scales will come from BBN affected by the
annihilation process.

Rehm and Jedamzik\cite{RJ98} considered annihilation immediately before
nucleosynthesis, at temperatures T = 80 keV -- 1 MeV.
Because of the much faster diffusion of neutrons and antineutrons
(as compared to
protons and antiprotons) the annihilation reduces the net neutron
number\cite{Steigman},
leading to underproduction of $\UHe$.  This sets a limit
$R <$ few \% to the amount of antimatter relative to matter
in domains of size $r_A \sim 1$ cm at $T$ = 100 GeV
($4\times10^6$ m at $T$ = 1 keV).

We extend these results to larger domain sizes, for which annihilation
occurs during or after nucleosynthesis.  
Since our results for the small domains and early annihilation
agree with Rehm and Jedamzik,
we concentrate on the larger domains and later annihilation in the
following discussion.  Below, all distance scales given in meters will
refer to comoving distance at $T = 1$ keV.

The case where annihilation occurs after nucleosynthesis was considered
in Refs.~\cite{He3ABBN}.
Because annihilation of antiprotons on helium would produce
D and $\EHe$ they estimated that the observed abundances of these
isotopes place an upper limit $R \lesssim 10^{-3}$ to the amount of
antimatter
annihilated after nucleosynthesis.  As we explain below,
the situation is actually more complicated.


Consider the evolution of an antimatter domain (of diameter $2r$)
surrounded by a larger region of matter.
At first matter and antimatter are in
the form of nucleons and antinucleons, after nucleosynthesis in the form
of ions and anti-ions.  Matter and antimatter will get mixed 
by diffusion and annihilated at the domain boundary.
Thus there will be a narrow
annihilation zone, with lower density, separating the matter and
antimatter domains.
At lower temperatures ($T <$ 30 keV)
the pressure gradient\cite{JF94} drives matter
and antimatter towards the annihilation zone.  This flow is resisted by
Thomson drag, which leads to diffusive flow\cite{CR98}.

Before nucleosynthesis, the mixing of matter and
antimatter is due to (anti)neutron diffusion.  When $\UHe$ is formed,
free neutrons disappear, and the annihilation practically ceases.
If annihilation is not complete by then, it is delayed
significantly because 
ion diffusion is much slower than neutron diffusion. 
There will then be a second burst of annihilation well after
nucleosynthesis, at $T \sim 1$ keV or below.
Indeed, depending on the size of the antimatter domains,
most of the annihilation occurs either
at $T > 80$ keV (for $r < 2\times10^7$ m)
or at $T < 3$ keV (for $r > 2\times10^7$ m).


The annihilation is so rapid that the outcome is
not sensitive to the annihilation cross sections.
The exact yields of the annihilation reactions are more important. 
{}From the Low-Energy Antiproton Ring (LEAR) at CERN,
we have data for antiprotons on helium,
and also for some other reactions with
antiprotons\cite{LEAR,Egidy,Balestra88}.  

The annihilation of a nucleon and an antinucleon produces a number of
pions, on average 5 with 3 of them charged\cite{Egidy}.
The charged pions decay into muons and neutrinos, the muons into
electrons and neutrinos.  The neutral pions decay into two photons.
About half of the annihilation energy, 1880 MeV, is carried away by the
neutrinos, one third by the photons, and one sixth by electrons and
positrons\cite{Steigman}.

If the annihilation occurs in a nucleus, some of the pions may knock out
other nucleons.
Part of the annihilation energy will go into the kinetic energy 
of these particles and the recoil energy of the residual nucleus. 
Experimental data on the energy spectra of these emitted nucleons
are well approximated by the formula $Ce^{-E/E_0}$, with average
energy $E_0 \sim 70$ MeV, corresponding to a momentum of
350 MeV/$c$\cite{Egidy}.

After $\UHe$ synthesis, the most important
annihilation reactions are $\barp\p$ and $\barp\UHe$.
According to Balestra \etal\cite{Balestra88}, a $\barp\UHe$ annihilation
leaves behind a $\EH$ nucleus in $43.7\pm3.2$ \% and a $\EHe$ nucleus in
$21.0\pm0.9$ \% of the cases.  The rms momentum of the residual
$\EHe$ was found to be $198\pm9$ MeV/$c$.

It is important
to consider how these annihilation products are slowed down.
If they escape far from
the antimatter domain, they will survive; but if they are thermalized
close to it they will soon be sucked into the annihilation
zone\cite{Aly}.

Fast ions lose
energy by Coulomb scattering on electrons and ions.
If the velocity of the ion is greater than thermal
electron velocities,
the energy loss is mainly due to 
electrons. At lower energies the scattering on ions becomes more
important.
Below $T = 30$ keV, when the thermal electron-positron pairs have
disappeared, the penetration distance of an ion of initial energy $E$
depends on the ratio $E/T$\cite{inpreparation}.
For $E \gg (M_{\rm ion}/m_e) T$,
the penetration distance is\cite{Jackson}
\beq
   l = \frac{m_e}{M_{\rm ion}}\frac{E^2}{4\pi n_e(Z\alpha)^2\Lambda}
   \approx 2\times10^9\frac{1}{AZ^2}\frac{1}{\eta_{10}}\frac{E^2}{T^3},
\label{pendist}
\eeq
where $\Lambda \sim 15$ is the Coulomb logarithm,
giving a comoving distance
\beq
   l_{\rm comoving} \approx
   \frac{1}{AZ^2}\frac{1}{\eta_{10}}\biggl(\frac{E}{T}\biggr)^2
   0.4\, {\rm m}.
\label{pendistc}
\eeq
For smaller $E/T$,  $l$ keeps getting shorter, but not as fast
as Eqs.~(\ref{pendist}) and (\ref{pendistc})
would give\cite{inpreparation}.

For $\EH$ and especially for $\EHe$, $l$ would become comparable to
the original size of the antimatter domain only well after the annihilation
is over.  Thus only a small fraction of these annihilation products escape
annihilation.  For D this fraction is larger, but still small, except for
the largest domains considered here.

Neutrons scatter on ions,
losing a substantial part of their energy in each collision.
The neutrons from annihilation reactions have
sufficient energy
to disintegrate a $\UHe$ nucleus.
This hadrodestruction\cite{DEHS} of $\UHe$ causes some additional $\EHe$ and 
D production.
Because protons are more abundant than
$\UHe$ nuclei, a neutron is more likely to scatter on a proton.
The mean free path
$\lambda=1/(\sigma_{np}n_p)$ is larger than the distance scales
considered here, so the annihilation neutrons are spread out evenly.
At lower temperatures ($T\lesssim1$ keV),
neutrons decay into protons before thermalizing.
At higher temperatures, the stopped
neutrons form deuterium with protons.


The high-energy photons and electrons from pion decay initiate 
electromagnetic cascades\cite{DEHS,photo,Ellis92,PSB95}.
Below $T = 30$ keV,
the dominant processes are photon-photon
pair production and inverse Compton scattering
\beq
   \gamma + \gamma_{bb} \rightarrow e^+ + e^-,\quad
   e + \gamma_{bb} \rightarrow e' + \gamma',
\eeq
with the background photons $\gamma_{bb}$.
The cascade photon energies $E_\gamma$ fall rapidly
until they are below the threshold 
for pair production,
$    
   E_\gamma \epsilon_\gamma = m_e^2,
$    
where $\epsilon_\gamma$ is the energy of the background photon.
Because of the large number of background photons, a significant 
number of them have energies $\gg T$, and the photon-photon pair production
is the dominant energy loss mechanism for cascade photons down 
to\cite{Ellis92}
\beq
   E_{\rm max} = \frac{m_e^2}{22T}.
\eeq
When the energy of a $\gamma$
falls below $E_{\rm max}$,
its mean free path increases and it is more likely to encounter an ion.

As the background temperature falls this threshold energy rises,
and below $T \sim 5$ keV, $E_{\rm max}$
becomes larger than nuclear binding energies,
and photodisintegration\cite{DEHS,photo,Ellis92,PSB95} becomes important.
Photodisintegration of D begins when the temperature falls below 5.3 keV,
photodisintegration of $\EHe$ ($\EH$) below 2.2 keV (1.9 keV) 
and photodisintegration of $\UHe$ below 0.6 keV.

Thus there are two regimes for photodisintegration:
(1) between $T = 5.3$ keV
and $T = 0.6$ keV, where the main effect is photodisintegration
of D, $\EH$, and $\EHe$; and 
(2) below $T = 0.6$ keV, where the main effect is production
of these lighter isotopes from $\UHe$ photodisintegration.
Because of the much
larger abundance of $\UHe$, even a small amount of annihilation
during the second regime swamps the effects
of the first regime, and only in the case that annihilation is
already over by $T = 0.6$ keV, is D photodisintegration
important\cite{DEHS,photo,Ellis92,PSB95}.
Because of the difference in the neutron and proton diffusion
rates, domains this small have already significant (neutron) annihilation
before $\UHe$ synthesis.

For the larger domain sizes, the most significant effect of
antimatter domains on BBN turns out to be $\EHe$ production
from $\UHe$ photodisintegration.


We have done numerical computations of nucleosynthesis with antimatter
domains.
Our inhomogeneous nucleosynthesis code
includes nuclear
reactions, diffusion, hydrodynamic expansion, annihilation, spreading
of annihilation products, photodisintegration of $\UHe$ and disintegration
by fast neutrons\cite{inpreparation}.  
Because of the lack
of data on the yields of annihilation reactions between nuclei and
antinuclei, we have not incorporated antinucleosynthesis in our code,
but the antimatter is allowed to remain as antinucleons.
This could affect our results at the 10\% level.

For photodisintegration of $\UHe$ we use the results of Protheroe
\etal\cite{PSB95} scaled by the actual local $\UHe$ abundance.
The $\EHe$ yield is an order of magnitude greater than the 
D yield\cite{PSB95}.
The Protheroe \etal\ results assume a standard
cascade spectrum.
This will not be
valid for temperatures below 100 eV,
since $E_{\rm max}$ becomes
comparable or greater than typical energies of the initial
$\gamma$'s from annihilation.  It would be important to find out the
true cascade spectrum for these low temperatures, since this will
affect our results at the largest scales.

We have in mind a situation where antimatter domains of typical diameter
$2r$ are separated by an average distance $2L$.
This we represent with a spherically symmetric grid of
radius $L$ with antimatter at the center with radius $r$.  For
simplicity we assume equal homogeneous initial densities for both the
matter and antimatter regions.  This density is set so that the final
average density after annihilation will correspond to a given
baryon-to-photon ratio $\eta$.  Since we are looking for upper limits
to the amount of antimatter which come from a lower limit to the
$\UHe$ abundance and an upper limit to the $\EHe$ abundance, we choose
$\eta = 6\times10^{-10}$, near the upper end of the acceptable range
in standard BBN, giving high $\UHe$ and low $\EHe$.

\begin{figure}[tbh]
\epsfysize=6.7cm
\epsffile{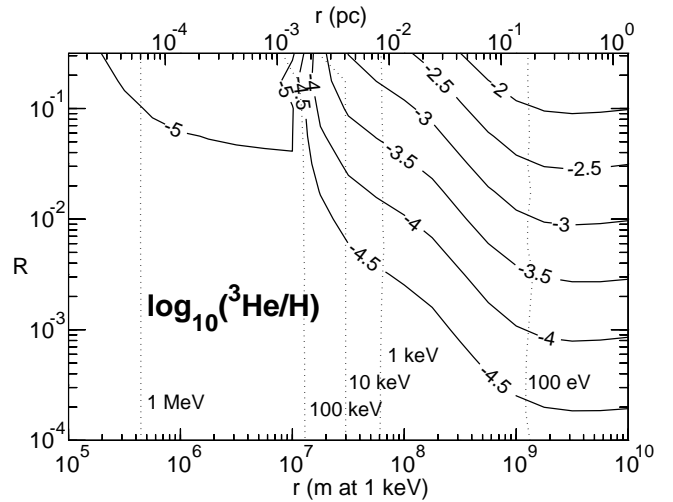}
\vspace*{0.3cm}
\caption[a]{\protect
The $\EHe$ yield as a function of
the antimatter/matter ratio $R$
and the antimatter domain radius $r$.  The distance
scales are given both at $T = 1$ keV (in meters)
and today (in parsecs).
We plot contours of (the logarithm of) the
number ratio $\EHe$/H.  The dotted lines show contours of the
``median annihilation temperature'', i.e., the temperature of the
universe when 50\% of the antimatter has annihilated.  Typically the
annihilation is complete at a temperature lower
than this
by about a factor of 3.
}
\label{fig:yields}
\bigskip
\end{figure}

We show the $\EHe$ yield as a function of the antimatter domain
radius $r$ 
and the antimatter/matter ratio $R$ in Fig.~\ref{fig:yields}.
For domains smaller than $r = 10^5$ m,
annihilation happens
before weak freeze-out, and has no effect on BBN.
For domains between $r = 10^5$ m and $r = 10^7$ m,
neutron annihilation before
$\UHe$ formation leads to a reduction in $\UHe$ and $\EHe$ yields.
For domains larger than $r = 2\times10^7$ m, most of the annihilation
happens after $\UHe$ synthesis.
Antiproton-helium annihilation then produces $\EHe$ and D, but most
of this is deposited close to the annihilation zone and is soon
annihilated.
The much more
important effect is the photodisintegration of $\UHe$ by the cascade
photons, since it takes place everywhere in
the matter region and thus the photodisintegration products
survive.  This leads to a large final $\EHe$ and D yield.
The
same applies to ${\rm n}\UHe$ reactions by fast neutrons from
$\barp\UHe$ annihilation, but the effect is much smaller,
because 
$\barp\UHe$ annihilation is less frequent than
$\barp\p$ annihilation, and a smaller part of the
annihilation energy goes into neutrons
than in the electromagnetic cascades.

\begin{figure}[tbh]
\epsfysize=6.7cm
\epsffile{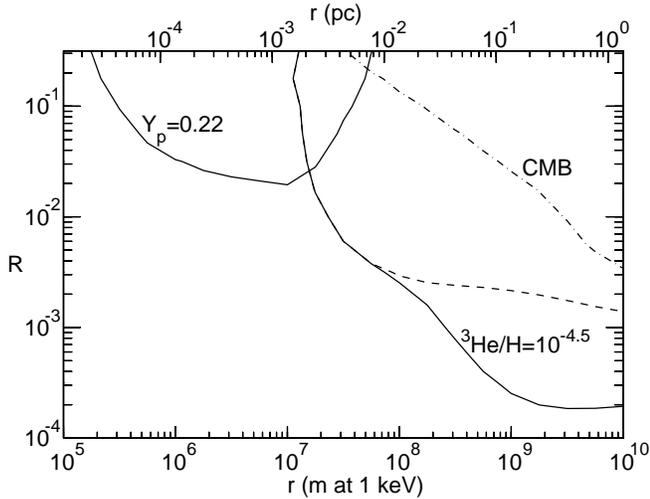}
\vspace*{0.3cm}
\caption[a]{\protect
Upper limits from BBN (solid lines) and CMB (dot-dashed line)
to the antimatter/matter ratio $R$
as a function of the antimatter domain radius $r$.
The dashed line gives the upper limit from BBN if 
photodisintegration is ignored.
}
\label{fig:limits}
\bigskip
\end{figure}

We obtain upper limits to the amount of antimatter in the early
universe by requiring that the primordial $\UHe$ abundance $Y_p$
must not be lower than $Y_p = 0.22$, and that the 
primordial $\EHe$ abundance must not be higher than
$\EHe$/H = $10^{-4.5}$\cite{BBNobs}.  
(The standard BBN results for $\eta = 6\times10^{-10}$ are
$Y_p = 0.248$ and $\EHe$/H = $1.1\times10^{-5}$.)
For domain sizes $r \lesssim 10^{11}$ m (or 10 parsecs today),
these limits are stronger
than those from the CMB spectrum distortion.
See Fig.~\ref{fig:limits}.

For $r < 10^5$ m, there is no BBN constraint on antimatter.
For $r = 10^5$--$10^7$ m, the amount of antimatter can be at most a few
per cent, to avoid $\UHe$ underproduction.
Our limit is somewhat weaker than that of Rehm and Jedamzik,
since they considered a lower $\eta = 3.4\times10^{-10}$.

For larger domains,
antimatter annihilation causes $\EHe$ production from $\UHe$
photodisintegration and the limit reaches
$R = 2\times10^{-4}$ at $r \sim 10^9$ m.

There may exist
small regions of parameter space where
acceptable light element yields
would be obtained for ``nonstandard''
values of $\eta$ and large $R$\cite{inpreparation}.
Clearly the simultaneous reduction of
$\UHe$ and increase of $\EHe$ and D suggest such a possibility
for large $\eta$.


We thank
T.~von Egidy, A.M.~Green, K.~Jedamzik, K.~Kajantie, 
J.~Rehm, J.-M.~Richard, M.~Sainio, G.~Steigman,
M.~Tosi, and S.~Wycech for useful discussions.
 We thank M.~Shaposhnikov for suggesting
this problem to us and P.~Ker\"{a}nen for reminding us about
photodisintegration.
We thank
the Center for Scientific Computing (Finland) for computational resources.

\end{document}